\documentstyle[12pt,epsf]{article}
\begin{document}
\baselineskip=20pt
                                             \begin{flushright}
                                             SPbU-IP-98-12 \\
                                             hep-ph/9805xxx
                                             \end{flushright}
\thispagestyle{empty}
\vskip 1.0cm

{\large
\centerline{
                Polarization Properties
                of Low Energy Amplitude
           }
\centerline{
                for
        $\pi N \to \pi \pi N$
                Reaction
           }
}

\vskip1.0cm
\centerline{ \bf
                A.A.~Bolokhov$ {}^{1} $,
                V.A.~Kozhevnikov$ {}^{1} $,
                S.G.~Sherman$ {}^{2} $ and
                D.N.~Tatarkin$ {}^{1} $
           }
\centerline{\sl
        $ {}^{1}  $
                Sankt-Petersburg State University,
           }
\centerline{\sl
                Sankt-Petersburg, 198904, Russia
           }

\centerline{ \sl
        $ {}^{2}  $
        St. Petersburg Institute for Nuclear Physics,
           }
\centerline{\sl
        Sankt--Petersburg, 188350, Russia
           }

\vskip 1.0cm

\centerline{\bf Abstract }
\vskip0.2cm

\noindent
                The theoretical study of cross sections
                for polarized--target measurements of
        $ \pi N \to \pi \pi N $
                reactions
                gives evidence
                that the interplay between the strong
                contribution from OPE mechanism
                and the one from isobar exchanges,
                which is equally strong within
                isobar half--widths energy region,
                must result in nontrivial polarization
                phenomena.
                The Monte--Carlo simulations for asymmetries
                in
        $ \pi^{-} p^{\uparrow} \to \pi^{-} \pi^{+} n $
                reaction at
        $ P_{\rm Lab} = 360 $~MeV/c
                with the use of theoretical amplitudes
                found as solutions for unpolarized data at
        $ P_{\rm Lab} < 500 $~MeV/c
                provide confirmations for significant effect.
                The effect is capable to discriminate between
                the OPE and isobar exchanges
                and it is sensitive to the OPE parameters
                in question.
                This leads to the conclusion
                that the decisive
        $\pi N \to \pi \pi N$
                analysis,
                aiming at determination of
        $ \pi \pi $--scattering
                lengths,
		must combine both unpolarized data
		and polarization information.
                The appropriate measurements are shown
                to be feasible at the already existing
                CHAOS spectrometer.

\vskip0.2cm

{
\noindent
          PACS number(s):
                13.75.-n,
                13.75.Gx,
                13.75.Lb
}

\vskip 1.5cm
\centerline{
                     Sankt--Petersburg }
\centerline{
                           1998 }

\newpage
\section{ Introduction }

                The
        $ \pi N \to \pi \pi N $
                reaction is considered to be
                an essential source of information
                on the
        $ \pi \pi $
                scattering.
                The values of
        $ \pi \pi $--scattering
                lengths
                can give restrictions for
                values of
                the effective low--energy parameters of QCD
                obtained within the framework of ChPT
                (Chiral Perturbation Theory)
                which was formulated by Gasser and Leutwiller
                in
\cite{GasserL82,GasserL8485}
                following Weinberg's ideas
\cite{Weinberg66,Weinberg68,Weinberg79}.
                The appearance of GChPT
                (Generalized ChPT)
                scheme
\cite{JSternSF93}
                enhanced
                the interest to the
        $ \pi \pi $
                interaction
                because of
                difference in the predicted
        $ \pi \pi $--scattering
                lengths
                with that given by ChPT.

                A review on the experimental opportunities
                for obtaining information on the
        $ \pi \pi $
                interaction
                and the discussion of
                the status of modern experiments
                planned to test the ChPT predictions
                might be found in the talks
\cite{Pocanic94,Pocanic98}
                by Po\v{c}ani\'{c}.
                The investigations of
        $ \pi N \to \pi \pi N $
                reactions are pronounced
                to be capable to
                discriminate between the ChPT and GChPT
                models for low--energy manifestations of QCD.
                Meanwhile
                the most recent attempts
\cite{BolokhovPS98zs1,PocanicF97,Olsson97,Patarakin97,Sevior97}
                of analysis of the
        $ \pi N \to \pi \pi N $
                data
                did not provide the necessary accuracy.

                One must recall
                that there are three methods for the
        $ \pi N \to \pi \pi N $
                data treatment
                applicable at the low energies.
                These are:

          {\bf 1.}
                The approach
                by Olsson and Turner
\cite{OlssonT686972},
                presenting the formulae
                that express the
        $ \pi \pi $--scattering
                lengths
        $ a^{I=0}_{0} $,
        $ a^{I=2}_{0} $
                in terms of the
                threshold characteristics of the
        $ \pi N \to \pi \pi N $
                reactions;

          {\bf 2.}
                The model--independent
                Chew--Low extrapolation procedure
                by Goebel, Chew and Low
\cite{ChewL};

          {\bf 3.}
                The method
                of determination of
                the OPE parameters directly
                in the physical region of the reaction
                by Oset and Vicente--Vacas
\cite{OsetV85};
                it is based on the
        $ \pi N \to \pi \pi N $
                phenomenological amplitude
                accounting for exchanges
                of the appropriate resonances.

                Due to simplicity of the  approach {\bf 1}
                and the statistical reliability of
                the total--cross--section data
                it operates with,
                the approach gained a broad scale
                of applications
                when the modern data started to appear
                in the close--to--threshold
                energy region
\cite{Kernl-0p89,Kernl+-n89,Kernl++n90,KernelEtAl91},
\cite{SeviorEtAl91},
\cite{Lowe_00n91},
\cite{PocanicEtAl94}.

                Recently
                the heavy baryon approximation
                was used to derive corrections
                to the Olsson--Turner formulae
\cite{BernardKM94,OlssonMKB95}.
                The progress of the later theoretical
                calculations of the
        $ \pi N \to \pi \pi N $
                amplitude
                within the same framework
\cite{BernardKM95,Meissner95,BernardKM97}
                resulted in conclusion
                that uncertainty in isobar interactions
                prevented from accurate determination
                of the isospin--zero scattering length
        $ a^{I=0}_{0} $.
                Finally,
                it was stated in
\cite{BernardKM97Trento}
                that the result for
        $ a^{I=0}_{0} $
                of the analysis
\cite{OlssonMKB95}
                cannot be trusted.

                The conclusion that
        $ a^{I=2}_{0} $
                acquired
                minor perturbations from isobars
                was made with the help of rather simple
                model of
        $ \pi \pi N N^{(*)} $
                and
        $ \pi \pi N \Delta $
                interactions.
                The investigations of various contributions
                to threshold amplitudes
                performed in the paper
\cite{BolokhovS98zs2}
                with more general isobar interactions
                raised doubts in the result for
        $ a^{I=2}_{0} $
                as well.
                Moreover,
                it was found
                that two independent threshold amplitudes for
        $ \pi N \to \pi \pi N $
                reactions
                collected noticeable contributions from the
        $ O (k^{4}) $
                parameters of the OPE graph.
                The determination of four
        $ \pi \pi $
                characteristics
                by virtue of only two
        $ \pi N \to \pi \pi N $
                threshold experimental quantities
                becomes a model--dependent problem
                that requires additional theoretical
                and experimental information.

                The most recent applications of the Chew--Low
                method {\bf 2} to the near--threshold data
                provided the
        $ \pi \pi $--scattering lengths
                with the almost perfect accuracy
\cite{PocanicF97,Patarakin97}.
                However,
                because of
                making use of
                assumptions about
        $ \pi N \to \pi \pi N $
                amplitude
                at zero momentum transfer,
                this method       suffers from unknown
                systematic errors
                which cannot be estimated within the approach.
		The known problems of applications of the
		method
                at higher energies are
		discussed in
\cite{Leksin70,MartinMS76}.
		The analysis of properties of the extrapolation
		procedure in the near--threshold region is made
		in
\cite{BolokhovS98zs2}
                in the framework of the approach {\bf 3}.
                The theoretical amplitudes were used to show
                that the strong influence of
		isobars in this region
                can result in inapplicability
                to synthetic data
                of both linear and
                quadratic extrapolations.
		The reported failure of the Chew--Low analysis
		of data for the channel
        $ \pi^{+} p \to \pi^{+} \pi^{+} n$
\cite{Patarakin97}
                might serve as an indirect evidence
                for disregard of other
                successful extrapolations
                since there might be observed a
                contradiction with the previous
                statement that the
        $ I_{\pi \pi} = 2 $
                channel is less
                perturbed near threshold by isobar
                contributions
\cite{BernardKM95,Meissner95,BernardKM97,BernardKM97Trento},
                see the previous paragraph;
                this definitely requires a separate explanation.
                The
                announced in the
                talk
\cite{Patarakin97}
                long awaited analysis
                of real data of the best quality obtained
                at the CHAOS device
\cite{SmithEtAl95}
                must become a checkpoint for low--energy
                applications of
                the discused method.

                The relativistic version of the approach
                {\bf 3}
                was used recently
\cite{BolokhovPS98zs1}
                to treat the
                large set of the near--threshold data
                on total cross sections and 1--D distributions
                in the energy region
        $ 300 \le P_{\rm Lab} \le 500 $~MeV/c.
                The phenomenological amplitude
                for the
                reaction
        $\pi N \to \pi \pi N$,
                taking into account the exchanges of
        $\Delta$
                and
        $N^{(*)}$
                along with the OPE mechanism
                and polynomial background
                derived with the account of
                isotopic, crossing,
        $ C $,
        $ P $
                and
        $ T $
                symmetries of strong interactions,
                was fitted
                to the
                experimental data.
                Though
                the parameters of OPE
                are found to be statistically significant,
                the
        $  \pi \pi $--scattering
                lengths
                appear different in different solutions.
                The origin of difficulties is attributed
                to the influence of isobars.
                Eight parameters of
        $ \pi \pi N \Delta $
                and
        $ \pi \pi N N^{(*)} $
                interactions,
                being only weakly constrained by the
                widths of decays
        $ \Delta \to \pi \pi N $,
        $ N^{(*)} \to \pi \pi N $,
                strongly correlate with the OPE parameters
                in question.

                The common difficulty of the discussed
                methods originates from limitations to data
                of unpolarized measurements
                which cannot discriminate
                between contributions to
                different spin--structures of the reaction
                amplitude.
                Up to now the known polarization measurements
                of the
        $\pi N \to \pi \pi N$
                reactions had been performed
                at considerably higher energies,
                for example,
                at 5.98 GeV/c and 11.85 GeV/c
\cite{Lesquenetal85}
                and at 17.2 GeV/c
\cite{Grayeretal74}.
                The analyses of the polarized data
\cite{Svec92},
\cite{Becker79}
                already proved such measurements
                to be detailed sources of information on the
        $\pi \pi$
                interaction.

                The main goals of the present paper is to
                elaborate the theoretical framework for
                treating the polarization measurements of
        $\pi N \to \pi \pi N$
                reactions at low energies and to find out
                the suitable observables.
                We pay a separate attention for a study
                of the principal possibility to perform
                polarized
        $\pi N \to \pi \pi N$
                experiments at the already existing
                CHAOS spectrometer
\cite{SmithEtAl95}.
                The primary goal is to ensure the solution
                of
                the urgent problem:
                the obtaining of the
        $  \pi \pi $--scattering
                characteristics
                with the help of the solid bank of available
        $\pi N \to \pi \pi N$
                data
                and simple polarization measurements added.

                We base upon the general properties
                of the
        $\pi N \to \pi \pi N$
                amplitude.
                It describes five charge channels in terms
                of only four isoscalar functions.
                Moreover,
                three of these functions are strongly
                restricted by crossing symmetry to only
                one independent function
                (see
\cite{BolokhovVS88,BolokhovVP88,BolokhovVS91}).
                To take advantage of intimate relations
                between various channels
                we prefer to rely upon
                the crossing symmetry
                rather than the partial--wave expansion.
                Obviously,
                the approach
        {\bf 3}
                is assumed.


                The paper is organized as follows.
                The content of sect. 2 reminds
                the structure of the
        $ \pi N \to \pi \pi N $
                amplitude.
                Sect. 3 provides expressions
                necessary for calculations
                of cross sections in experiments
                with the polarized target.
                Sect. 4 is devoted to
                geometry of devices and
                analyzed asymmetries.
                It contains
                the results of modelling
                the experimental measurements
                with the use of various solutions
                found in the paper
\cite{BolokhovPS98zs1}
                for the
        $ \pi N \to \pi \pi N $
                amplitude.
                The summary, the concluding remarks
                and the discussion of
                implementations
                are given in Conclusions.

%
%

%


\section{
                General Structure of
            $\pi N \to \pi \pi N$
               Amplitude
}

                This short section introduces
                the basic formulae of
                the papers
\cite{BolokhovVS91,BolokhovPS98zs1}.

                We consider the reaction
\begin{equation}
\label{React}
                \pi^{a}(k_1) + N_{\alpha}(p;\lambda_i) \
                        \rightarrow   \
                        \pi^{b}(k_2) + \pi^{c}(k_3)
                        + N_{\beta}(q;\lambda_f)  \ ,
\end{equation}
                where
        $ a, b, c = 1, 2, 3$
                and
        $ \alpha, \beta = 1, 2$
                are isotopic indices of pions and nucleons,
                respectively,
                and
        $ \lambda_{i} $
        ($ \lambda_{f} $)
                are polarizations of
                initial
                (final)
                nucleons.

                Separating the nucleon spinor wave functions
                from the reaction amplitude
        $M^{abc}_{\beta \alpha} ( \lambda_f ; \lambda_i )$
\begin{equation}
\label{SAmp}
                M^{abc}_{\beta \alpha} ( \lambda_f ; \lambda_i )
                \ = \ \bar{u}(q;\lambda_f)
                \hat{M}^{abc}_{\beta \alpha} (i \gamma_5)
                {u}(p;\lambda_i) \ ,
\end{equation}
                one can define the isoscalar amplitudes
        $ \hat{A} $,
        $ \hat{B} $,
        $ \hat{C} $,
        $ \hat{D} $
                by
\begin{equation}
\label{IDec}
                \hat{M}^{abc}_{\beta \alpha} \ = \
                \hat{A} \tau^{a}_{\beta \alpha} \delta^{b c} +
                \hat{B} \tau^{b}_{\beta \alpha} \delta^{a c} +
                \hat{C} \tau^{c}_{\beta \alpha} \delta^{a b} +
                \hat{D} i \epsilon^{abc} \delta_{\beta \alpha} \ ,
\end{equation}
        $ \tau^{a} $,
        $ a = 1, 2, 3 $
                being the nucleon--isospin generators.
                The amplitudes of five observable channels
                are related to
                $ A, B, C, D $
                by
\begin{eqnarray}
\nonumber
                \hat{M}_{\{\pi^{-} p \to \pi^{-} \pi^{+} n\}} \ = \
                        \sqrt{2} / 2 \ (\hat{A} + \hat{C}) \ ,
        \; \;
                \hat{M}_{\{\pi^{-} p \to \pi^{0} \pi^{0} n\}} \ = \
                       \ \ 1 / 2 \ (\hat{A}) \ ,
        \; \;
\\
\nonumber
                \hat{M}_{\{\pi^{-} p \to \pi^{-} \pi^{0} p\}} \ = \
                        1 / 2 \ (\hat{C} - 2 \hat{D}) \ ,
        \; \;
                \hat{M}_{\{\pi^{+} p \to \pi^{+} \pi^{0} p\}} \ = \
                        1 / 2 \ (\hat{C} + 2 \hat{D}) \ ,
        \; \;
\\
\label{ChAmp}
                \hat{M}_{\{\pi^{+} p \to \pi^{+} \pi^{+} n\}} \ = \
                        1 / 2 \ (\hat{B} + \hat{C}) \ .
\end{eqnarray}
                To simplify the processing of cross sections,
                the statistical factors
                accounting for identical pions
                are inserted into these definitions.
                Below, we leave only the charges of the
                final pions in subscripts for the channels.

                To decompose
                each isoscalar function
        $ \hat{A}, \hat{B}, \hat{C}, \hat{D} $
                and each amplitude
        $ \hat{M}_{X} $,
        $ X =
              \{-+n\}, $
        $      \{-0p\}, $
        $      \{00n\}, $
        $      \{++n\}, $
        $      \{+0p\}  $
                into
                independent spinor form factors
                let us define
                the crossing--covariant
                complex
                combinations
        ${k} = k_{R} + i k_{I} $,
        $ \bar{k} = k_{R} - i k_{I} $
                of pion momenta
\begin{eqnarray}
\label{vark}
                k =  - k_1 + \epsilon k_2 + \bar{\epsilon} k_3  \ ,
        \; \; \;
                \bar{k} = - k_1 + \bar{\epsilon} k_2 + \epsilon k_3  \ ,
\\
                k_{R} =  - k_1 - ( k_2 + k_3 ) / 2 \ , \; \;
                k_{I} =  \sqrt{3} ( k_2 - k_3 ) / 2 \ ,
\end{eqnarray}
                where
     $ \epsilon \equiv \exp (2 \pi i / 3 )
                \ = \ - 1 / 2 + i \sqrt{3} / 2 $,
     $ \bar{\epsilon} = \epsilon^{*} \ = \ - 1 / 2 - i \sqrt{3} / 2 $.
                The decomposition reads
\begin{eqnarray}
\label{DecAmp}
                \hat{M}_{X}
                       & \ = \ &
                        S_X +
                        \bar{V}_X \hat{k} +
                        {V}_X \hat{\bar{k}} +
                        i/2 \ T_X [ \hat{k} , \hat{\bar{k}} ]
                 \equiv
        \left(
\begin{array}{c}
                S_{X}^{}
\\
                \bar{V}_{X}^{}
\\
                V_{X}^{}
\\
                T_{X}^{}
\end{array}
        \right)^{\rm T}
                        \cdot
        \left(
\begin{array}{c}
                \hat{1}
\\
                \hat{k}
\\
                \hat{\bar{k}}

\\
                i/2
                [
                \hat{k},
                \hat{\bar{k}}
                ]
\end{array}
        \right)
                \ ,
                \; \;
\\
\nonumber
        &&
         (X =
              \{-+n\},
              \{-0p\},
              \{00n\},
              \{++n\},
              \{+0p\})
                \ .
\end{eqnarray}

                Since polarization phenomena
                are determined by the interference
                of real and imaginary parts
                of the amplitude,
                it is convenient to deal with combinations
\begin{equation}
\label{SVRVIT}
                S_{X}
                \; , \; \;
                V^{R}_{X} \equiv ( V_{X} + \bar{V}_X ) / 2
                \; , \; \;
                V^{I}_{X} \equiv ( V_{X} - \bar{V}_X ) / (2i)
                \; , \; \;
                T_{X}
                \; ,
\end{equation}
                which are shown to be approximately real
                in the energy region
                where unitarity corrections are small
                (see
\cite{BolokhovVP88}).
                So we rewrite the decomposition
(\ref{DecAmp})
                in the form:
\begin{eqnarray}
\label{DecAmpR}
                \hat{M}_{X}
                       & \ = \ &
        \left(
\begin{array}{c}
                S_{X}^{}
\\
                {V}^{R}_{X}
\\
                V^{I}_{X}
\\
                T_{X}^{}
\end{array}
        \right)^{\rm T}
                        \cdot
        \left(
\begin{array}{c}
                \hat{1}
\\
                2 \hat{k}_{R}
\\
                2 \hat{k}_{I}
\\
                {}
                [
                \hat{k}_{R},
                \hat{k}_{I}
                ]
\end{array}
        \right)
                \ ,
                \; \;
\\
\nonumber
        &&
         (X =
              \{-+n\},
              \{-0p\},
              \{00n\},
              \{++n\},
              \{+0p\})
                \ .
\end{eqnarray}

                The matrix element
        $ ||M||^{2} $
                entering the unpolarized cross section
                is the sum over final polarizations and
                the average over initial ones.
                It is the quadratic form of the
                vector of spinor form factors
        $ ( S_{X}, V^{R}_X, V^{I}_{X}, T_{X} ) $:
\begin{eqnarray}
\nonumber
                {\| M_{X} \|}^2
        & \equiv &
                1/2 \ \sum_{\lambda_f, \lambda_i}
                  \left[
                           \bar{u}(q;\lambda_f) \hat{M}_{X}
                           (i \gamma_5) {u}(p;\lambda_i)
                  \right]
                                \;
                  \left[
                           \bar{u}(q;\lambda_f) \hat{M}_{X}
                           (i \gamma_5) {u}(p;\lambda_i)
                  \right]^{*}
\\
\label{SqAmpR}
        & = &
        \left(
\begin{array}{c}
                S_{X}
\\
                V^{R}_{X}
\\
                V^{I}_{X}
\\
                T_{X}
\end{array}
        \right)^{\rm T}
                          G_{R}
        \left(
\begin{array}{c}
                S_{X}
\\
                V^{R}_{X}
\\
                V^{I}_{X}
\\
                T_{X}
\end{array}
        \right)^{*}
         =
        \left(
\begin{array}{c}
                S_{X}
\\
                V^{R}_{X}
\\
                V^{I}_{X}
\\
                T_{X}
\end{array}
        \right)^{\dagger}
                          G_{R}
        \left(
\begin{array}{c}
                S_{X}
\\
                V^{R}_{X}
\\
                V^{I}_{X}
\\
                T_{X}
\end{array}
        \right)
                           \ , \;
\\
\nonumber
        &&
             (X=\{-+n\},\{-0 \ p\},\{0 \ 0 \ n\},
                                \{++n\},\{+0 \ p\}) \ .
\end{eqnarray}
                The real hermitian matrix
        $ G_{R} $
                is obtained by calculating the
        $ \gamma $--matrix
                traces
\begin{eqnarray}
\label{GmatR}
                G_{R}
        & \equiv &
                \frac{1}{2}
                        {\rm Sp}
                        \left[
                          (\hat{q} + m)
        \left\{
\begin{array}{c}
                \hat{1}
\\
                2 \hat{k}_{R}
\\
                2 \hat{k}_{I}
\\
                {}
                [
                \hat{k}_{R},
                \hat{k}_{I}
                ]
\end{array}
        \right\}
                          (\hat{p} - m)
                        \gamma_{0}
        \left\{
\begin{array}{c}
                \hat{1}
\\
                2 \hat{k}_{R}
\\
                2 \hat{k}_{I}
\\
                {}
                [
                \hat{k}_{R},
                \hat{k}_{I}
                ]
\end{array}
        \right\}^{\dagger}
                        \gamma_{0}
                        \right]
                           \ .
\end{eqnarray}
                Its explicit expression will be given below,
                see eqs.
(\ref{GRmat}).

\section{
                Cross Section
                for Polarized--Target Measurements
\label{xsectPT}
        }


                The origin of strong correlations between
                parameters of OPE and isobar contributions,
                preventing from the accurate determination of
                the
        $  \pi \pi $--scattering
                lengths in the unpolarized experiment,
                is obvious now.
                Only the specific combination of the competing
                contributions given by eq.
(\ref{SqAmpR})
                can be measured in such experiments.
                Bringing this matrix to the diagonal form
                one can realize
                that any
                diagonal amplitude can mimic
                the OPE one outside the region
                of isobar poles.

                Though the measurement
                of the final polarization in the
        $\pi N \to \pi \pi N$
                reaction is implied by the design of
                the spectrometer AMPIR
                (see
\cite{KurepinEtAl92}),
                such measurements are hardly to be performed
                in the near future.
                Therefore,
                we consider the
                polarized--target
                experimental setup.
                For simplicity,
                we assume the ideal polarization.
                It is easy to generalize our results
                to the incomplete polarization
                due to the linear dependence
                of all asymmetries
                upon the polarization vector
        $ {\bf s} $.
                Indeed,
                given nontrivial probabilities
        $ w_{\lambda_{1}} $,
        $ w_{\lambda_{2}} $
        ($ w_{\lambda_{1}} + w_{\lambda_{2}} = 1 $)
                for the projection of the initial nucleon
                spin in the direction
        $ {\bf n}_{\bf s} \equiv
         {\bf s} / |{\bf s}| $
                to be
        $ \lambda_{1} = 1/2 $,
        $ \lambda_{2} = -1/2 $,
                respectively,
                any theoretical result for asymmetry
                must be derived with
        $ |{\bf s}| = w_{\bf s} = 2 w_{\lambda_{1}} - 1
                = w_{\lambda_{1}} - w_{\lambda_{2}} $.
                We set
        $ w_{\bf s} = 1  $
                in calculations.

                The matrix element
        $ \| M \|^{2}_{\bf s} $
                is now defined by
\begin{eqnarray}
\nonumber
                {\| M_{X} \|}^2_{\bf s}
        & \equiv &
                 \ \sum_{\lambda_f }
                  \left[
                           \bar{u}(q;\lambda_f) \hat{M}_{X}
                           (i \gamma_5) {u}(p;\lambda_i)
                  \right]
                                \;
                  \left[
                           \bar{u}(q;\lambda_f) \hat{M}_{X}
                           (i \gamma_5) {u}(p;\lambda_i)
                  \right]^{*}
\\
\label{SqAmps}
        & = &
        \left(
\begin{array}{c}
                S_{X}
\\
                V^{R}_{X}
\\
                V^{I}_{X}
\\
                T_{X}
\end{array}
        \right)^{\rm T}
                          G
        \left(
\begin{array}{c}
                S_{X}
\\
                V^{R}_{X}
\\
                V^{I}_{X}
\\
                T_{X}
\end{array}
        \right)^{*}
         =
        \left(
\begin{array}{c}
                S_{X}
\\
                V^{R}_{X}
\\
                V^{I}_{X}
\\
                T_{X}
\end{array}
        \right)^{\dagger}
                          G^{*}
        \left(
\begin{array}{c}
                S_{X}
\\
                V^{R}_{X}
\\
                V^{I}_{X}
\\
                T_{X}
\end{array}
        \right)
                           \ , \;
\\
\nonumber
        &&
             (X=\{-+n\},\{-0 \ p\},\{0 \ 0 \ n\},
                                \{++n\},\{+0 \ p\}) \ .
\end{eqnarray}
                The hermitian matrix
        $ G \equiv G_{R} + i G_{I} $
                is given by
\begin{eqnarray}
\label{Gmat}
                G
        & \equiv &
                        {\rm Sp}
                        \left[
                          (\hat{q} + m_{f})
        \left\{
\begin{array}{c}
                \hat{1}
\\
                2 \hat{k}_{R}
\\
                2 \hat{k}_{I}
\\
                {}
                [
                \hat{k}_{R},
                \hat{k}_{I}
                ]
\end{array}
        \right\}
                          (\hat{p} - m_{i})
                        \frac
                        { 1 + \gamma_{5} \hat{s} }
                        {2}
                        \gamma_{0}
        \left\{
\begin{array}{c}
                \hat{1}
\\
                2 \hat{k}_{R}
\\
                2 \hat{k}_{I}
\\
                {}
                [
                \hat{k}_{R},
                \hat{k}_{I}
                ]
\end{array}
        \right\}^{\dagger}
                        \gamma_{0}
                        \right]
                           \ ,
\end{eqnarray}
                where the polarization 4--vector
        $ s $
                equals
        $ ( 0, {\bf s} ) $
                in the rest frame of the initial nucleon.

                The real part
        $ G_{R} $
                of this matrix enters the unpolarized
                cross section
                (conf. eqs.
(\ref{GmatR})
                and
(\ref{Gmat})).
                The imaginary part
        $ G_{I} $
                is skew--symmetric.
                These matrices are explicitly given by
\begin{eqnarray}
\nonumber
      G_R (1,1) & = & 2 \, ( - m_i m_f + p \cdot q )
     \; ,
\\
\nonumber
      G_R (1,2) & = & 4 \, ( - m_i \, q \cdot k_R + m_f \, p \cdot k_R)
     \; ,
\\
\nonumber
      G_R (1,3) & = & 4 \, ( - m_i \, q \cdot k_I + m_f \, p \cdot k_I)
     \; ,
\\
\nonumber
      G_R (1,4) & = & 4 \, ( - p \cdot k_R \, q \cdot k_I
                             + q \cdot k_R \, p \cdot k_I)
     \; ,
\\
\nonumber
      G_R (2,2) & = & 8 \, ( - m_i m_f \, k_R \cdot k_R
                             - p \cdot q \, k_R \cdot k_R
                             + 2 \, p \cdot k_R \, q \cdot k_R)
     \; ,
\\
\nonumber
      G_R (2,3) & = & 8 \, ( - m_i m_f \, k_R \cdot k_I
                             - p \cdot q \, k_R \cdot k_I
                             + p \cdot k_R \, q \cdot k_I
\\
\nonumber
        &&
                            + q \cdot k_R \, p \cdot k_I)                                \; ,
\\
\nonumber
      G_R (2,4) & = & 8 \, ( - m_i \, q \cdot k_R \, k_R \cdot k_I
                             + m_i \, q \cdot k_I \, k_R \cdot k_R
\\
\nonumber
        &&
                             - m_f \, p \cdot k_R \, k_R \cdot k_I
                             + m_f \, p \cdot k_I \, k_R \cdot k_R)
     \; ,
\\
\nonumber
      G_R (3,3) & = & 8 \, ( - m_i m_f \, k_I \cdot k_I
                             - p \cdot q \, k_I \cdot k_I
                             + 2 \, p \cdot k_I \, q \cdot k_I)
     \; ,
\\
\nonumber
      G_R (3,4) & = & 8 \, ( - m_i \, q \cdot k_R \, k_I \cdot k_I
                             + m_i \, q \cdot k_I \, k_R \cdot k_I
\\
\nonumber
        &&
                             - m_f \, p \cdot k_R \, k_I \cdot k_I
                             + m_f \, p \cdot k_I \, k_R \cdot k_I)
     \; ,
\\
\nonumber
      G_R (4,4) & = & 8 \, ( - m_i m_f \, k_R \cdot k_R \, k_I \cdot k_I
                             + m_i m_f \, (k_R \cdot k_I)^{2}
\\
\nonumber
        &&
                             + p \cdot q \, k_R \cdot k_R \, k_I \cdot k_I
                             - p \cdot q \, (k_R \cdot k_I)^{2}
\\
\nonumber
        &&
                  - 2 \, p \cdot k_R \, q \cdot k_R \, k_I \cdot k_I
                  + 2 \, p \cdot k_R \, q \cdot k_I \, k_R \cdot k_I
\\
\label{GRmat}
        &&
                  + 2 \, q \cdot k_R \, p \cdot k_I \, k_R \cdot k_I
                  - 2 \, p \cdot k_I \, q \cdot k_I \, k_R \cdot k_R)
     \; ,
\end{eqnarray}
\begin{eqnarray}
\nonumber
      G_I (1,2) & = & - 4 \, {\rm eps} [p,q,s,k_R]
     \; ,
\\
\nonumber
      G_I (1,3) & = & - 4 \, {\rm eps} [p,q,s,k_I]
     \; ,
\\
\nonumber
      G_I (1,4) & = & 4 \, ( {\rm eps} [p,s,k_R,k_I] m_f
                  - {\rm eps} [q,s,k_R,k_I] m_i)
     \; ,
\\
\nonumber
      G_I (2,3) & = &   8 \, ({\rm eps} [p,s,k_R,k_I] m_f
                  + {\rm eps} [q,s,k_R,k_I] m_i)
     \; ,
\\
\nonumber
      G_I (2,4) & = & 8 \, (  {\rm eps} [p,q,s,k_R] \, k_R \cdot k_I
                  - {\rm eps} [p,q,s,k_I] \, k_R \cdot k_R
\\
\nonumber
        &&
                  + 2 \, {\rm eps} [p,q,k_R,k_I] \, s \cdot k_R
                  + 2 \, {\rm eps} [q,s,k_R,k_I] \, p \cdot k_R)
    \; ,
\\
\nonumber
      G_I (3,4) & = & 8 \, (  {\rm eps} [p,q,s,k_R] \, k_I \cdot k_I
                  - {\rm eps} [p,q,s,k_I] \, k_R \cdot k_I
\\
\label{GImat}
        &&
                  + 2 \, {\rm eps} [p,q,k_R,k_I] \, s \cdot k_I
                  + 2 \, {\rm eps} [q,s,k_R,k_I] \, p \cdot k_I)
    \; ,
\end{eqnarray}
                where the following notation
        $$
      {\rm eps} [x,y,u,v]
         \equiv
                  \epsilon_{\mu \nu \rho \sigma}
                   x^{\mu} y^{\nu} u^{\rho} v^{\sigma}
        $$
                is used
                and nucleon masses
        $ m_{i} $,
        $ m_{f} $
                are allowed to be different.
                Matrix elements in eqs.
(\ref{GImat})
                are actually ordered according to
                their importance
                in the near--threshold region.

                Let us now consider the fixed reaction channel
                and omit the channel's subscript
                in the notation for the vector
        $ M $
                of form factors
(\ref{SVRVIT}).
                The form factors
        $ V^{R} $,
        $ V^{I} $
                are obtained by splitting off the real
                and imaginary parts in
                the cross--covariant momenta
                of the spinor structures
        $ \hat{k} $,
        $ \hat{\bar{k}} $.
                The form factors themselves remain
                complex:
        $ V^{R} \equiv V^{R}_{R} + i V^{R}_{I} $,
        $ V^{I} \equiv V^{I}_{R} + i V^{I}_{I} $.
                Consider
                the real and imaginary parts of
                the amplitude:
\begin{eqnarray}
\label{MRMI}
                  M_{R}
        & \equiv &
        \left(
\begin{array}{c}
                S_{R}
\\
                V^{R}_{R}
\\
                V^{I}_{R}
\\
                T_{R}
\end{array}
        \right)
        \; , \; \;
                  M_{I}
            \, \equiv \,
        \left(
\begin{array}{c}
                S_{I}
\\
                V^{R}_{I}
\\
                V^{I}_{I}
\\
                T_{I}
\end{array}
        \right)
        \; .
\end{eqnarray}
                Then the matrix element
(\ref{SqAmps})
                can be rewritten as
\begin{eqnarray}
\label{SqAmpsRI}
                {\| M \|}^2_{\bf s}
        & = &
                M^{\rm T}_{R} G_{R} M_{R} +
                M^{\rm T}_{I} G_{R} M_{I} +
               2 M^{\rm T}_{R} G_{I} M_{I}
                \; .
\end{eqnarray}
                Here,
                the first two terms of the right--hand side
                give the unpolarized matrix element
(\ref{SqAmpR}).
                The effect of polarization is provided by
                the third term.

                Two conclusions can be immediately derived
                from this form
                and the above explicit expressions
                for matrices
        $ G_{R} $,
        $ G_{I} $.

                {\bf 1.}
                OPE contributes only to the spinor form factor
        $ S $
                of the decompositions
(\ref{DecAmp}),
(\ref{DecAmpR}).
                Hence,
                the validity of the assumption about
                the OPE dominance means
                that there cannot be any asymmetry
                in the reaction cross sections
                at the energies
                where the assumption holds.

                {\bf 2.}
                It is necessary for polarization effect
                that both real and imaginary
                parts of the amplitude
                remain
                non--negligible.
                Fortunately,
                several partial waives are
                mixed up  in
                the polarization term of eq.
(\ref{SqAmpsRI}).
                So the polarization effect
                must manifest itself in
                asymmetries
                of cross sections not only
                at the isobar poles
        ($ P_{\rm Lab} \approx 500 $~Mev/c
                for
        $ \Delta $
                and
        $ P_{\rm Lab} \approx 660 $~Mev/c
                for
        $ N^{(*)} $)
                but
                well below due to the large
                widths of these resonances.

                Rich kinematics of the considered reaction
                give rise to the abundance of possibilities
                for manifestation of polarization
                in the polarized--target experiments.
                There are five distinct structures
                entering the matrix
        $ G_{I} $
                of eq.
(\ref{SqAmpsRI}).
                These are
\begin{eqnarray}
\label{epspqskR}
         {\rm eps} [p,q,s,k_R]
        & = &
         -3/2 \, {\rm eps} [p,q,s,k_1]
        \; ,
\\
\label{epspqskI}
         {\rm eps} [p,q,s,k_I]
        & , &
\\
\label{epspskRkI}
         {\rm eps} [p,s,k_R,k_I]
        & , &
\\
\label{epsqskRkI}
         {\rm eps} [q,s,k_R,k_I]
        & , &
\\
\label{epspqkRkI}
         {\rm eps} [p,q,k_R,k_I]
        & = &
        -3 \sqrt{3}/4 \, {\rm eps} [p,q,k_1,k_2 - k_3]
         \; .
\end{eqnarray}
                One finds that,
                depending on the relative strength
                of form factors
(\ref{MRMI}),
                any of the above structures
                can govern the discussed effects.
                Four structures
(\ref{epspqskR}),
(\ref{epspqskI}),
(\ref{epspskRkI}),
(\ref{epsqskRkI}),
                entering matrix elements
        $ G_{I} (1,2) $,
        $ G_{I} (1,3) $
                and
        $ G_{I} (1,4) $
                related to OPE,
                are also present in the rest elements
                given by eqs.
(\ref{GImat}).
                At small
        $ {\bf q} $,
        $ {\bf k}_{2} $,
        $ {\bf k}_{3} $,
                the terms
(\ref{epspqskR}),
(\ref{epspqskI})
                are the most favourable ones
                for detecting OPE
                since extra factors at the same terms in
        $ G_{I} (2,3) $,
        $ G_{I} (2,4) $,
        $ G_{I} (3,4) $
                eliminate the effect when averaged.
                At the same time,
                there is the single term
(\ref{epspqkRkI})
                which is
                specific to non--OPE contributions only.
                It can be ``switched off'' by
        $ (s \cdot k_{R}) $
                factor
                since vector
        $ {\bf k}_{R} $
                belongs to narrow backward cone
                at low energies.
                This phenomenon is a characteristic feature
                of non--OPE mechanisms.


                Let two vectors
        $ {\bf x } $, $ {\bf y } $
                determine the plane
        $ ( {\bf x, y } ) $,
                separating ``left'' and
                ``right'' semi--spheres
                and
        $ {\bf z } $
                be some third vector.
                Let
        $ \varphi_{\bf z} $
                be its azimuthal angle
                in the plane which contains
        $ {\bf y} $
                and is orthogonal to the plane
        $ ( {\bf x, y } ) $.
                The asymmetry
\begin{eqnarray}
\label{Asym}
                A_{ ({\bf x, y}) } ( \varphi_{\bf z} )
        & \equiv &
                \frac{
                        \sigma(\varphi_{\bf z}) - \sigma(-\varphi_{\bf z})
                     }
                     {
                        \sigma(\varphi_{\bf z}) + \sigma(-\varphi_{\bf z})
                     }
                \;
\end{eqnarray}
                shows the relative value of the
                polarization term of eq.
(\ref{SqAmpsRI})
                with respect to unpolarized cross section.
                Obviously,
                several asymmetries must be observed
                to detect the influence of all above
                structures
(\ref{epspqskR})--(\ref{epspqkRkI}).

                It was already pointed out
                in the beginning of this section
                that the above formulae
(\ref{SqAmps}),
(\ref{Gmat}),
(\ref{GRmat}),
(\ref{GImat})
                and
(\ref{SqAmpsRI})
                remain valid for incomplete polarization
                of the target,
                the vector
        $ s \equiv s_{i} $
                being recognized as the polarization vector
                of the density matrix
        $ \rho_{i} $
                for the initial nucleon.
                The density matrix for the final nucleon
        $ \rho_{f} \equiv \frac{1}{2}
                          ( \hat{q} + m_{f} )
                          ( 1 - \gamma_{5} \hat{s}_{f} )$
                is given by
\begin{eqnarray}
\label{densf}
        \rho_{f}
                & = &
                \frac{
                        ( \hat{q} + m_{f} )
                           \hat{M} ( i \gamma5 ) \rho_{i}
                        \gamma_{0}
                                ( i \gamma5 )^{\dagger} \hat{M}^{\dagger}
                        \gamma_{0}
                        ( \hat{q} + m_{f} )
                     }
                     {
                        {\rm Sp}
                        \left[
                        ( \hat{q} + m_{f} )
                           \hat{M} ( i \gamma5 ) \rho_{i}
                        \gamma_{0}
                                ( i \gamma5 )^{\dagger} \hat{M}^{\dagger}
                        \gamma_{0}
                        \right]
                     }
                \; .
\end{eqnarray}
                The expression for the polarization vector
        $ s_{f} $
                takes the form
\begin{eqnarray}
\label{sf}
          s^{\mu}_{f}
                & = &
        \frac{ 1 }{ 2 m_{f} }
    \,
        \frac{
%
                M^{\rm T}
                          F^{\mu}
                M^{*}
%
         }
         {
%
                M^{\rm T}
                          G
                M^{*}
         }
        \; ,
\end{eqnarray}
                where
        $ G $
                is given by eqs.
(\ref{Gmat}),
(\ref{GRmat}),
(\ref{GImat})
                and the array of matrices
        $ F^{\mu} $
                can be calculated as
\begin{eqnarray}
\label{Fmat}
                F^{\mu}
         \equiv
                        {\rm Sp}
                        \left[
                          (\hat{q} + m_{f})
        \left\{
\begin{array}{c}
                \hat{1}
\\
                2 \hat{k}_{R}
\\
                2 \hat{k}_{I}
\\
                {}
                [
                \hat{k}_{R},
                \hat{k}_{I}
                ]
\end{array}
        \right\}
                          (\hat{p} - m_{i})
                        \frac
                        { 1 + \gamma_{5} \hat{s}_{i} }
                        {2}
                        \gamma_{0}
        \left\{
\begin{array}{c}
                \hat{1}
\\
                2 \hat{k}_{R}
\\
                2 \hat{k}_{I}
\\
                {}
                [
                \hat{k}_{R},
                \hat{k}_{I}
                ]
\end{array}
        \right\}^{\dagger}
                        \gamma_{0}
                          (\hat{q} + m_{f})
                        \gamma_{5}
                        \gamma^{\mu}
                        \right]
                           \, .
\end{eqnarray}
                The calculation with the use of the standard
                High Energy Physics package of computer
                algebra
\cite{Hern85}
                is straightforward,
                the result being too cumbersome to be displayed
                here.

                Values of
        $  s_{f} $
                and
        $ A_{ ({\bf x, y}) } ( \varphi_{\bf z} ) $
                are defined over 4--dimensional phase space
                of the considered reaction.
                This makes it difficult to display
                such quantities visually.
                Below,
                we consider asymmetries which are
                integrated as over
                ``orange lobules'' of
        $ \varphi_{\bf z} $
                bins,
        $ {\bf z} $
                being a selected momentum,
                as well as over allowed range of the rest
                momenta.
                This averaging
                suppresses
                the polarization effect.
                The suppression depends upon the kinematical
                symmetry of the considered amplitude:
                the more symmetry is displayed
                by the amplitude,
                the less value of the averaged asymmetry
        $ A_{ ({\bf x, y}) } ( \varphi_{\bf z} ) $
                is obtained.
                It was shown in the paper
\cite{BolokhovVP88}
                that form factors
        $ S $,
        $ V^{R} $,
        $ V^{I} $,
        $ T $
                of isoscalar amplitudes
        $ A $,
        $ B $,
        $ C $,
        $ D $
                had definite properties under permutation
                of nucleons
        $ p \leftrightarrow -q $
                due to charge--conjugation symmetry,
                the properties of
        $ D $--amplitude
                form factors being opposite to the ones
                of the corresponding form factors of the
                rest isoscalar amplitudes.
                Another symmetry of particle momenta,
                that eliminates the kinematical degrees of
                freedom,
                is related to Bose statistics of identical
                pions in
        $ \{++n\} $
                and
        $  \{0 \ 0 \ n\}$
                channels.
                The expressions for channel amplitudes
                given by eqs.
\ref{ChAmp}
                show that the asymmetries for the above two
                channels are more suffering from the discussed
                degeneracies
                while channels
        $  \{ \pm  0 \ p \} $,
                being free from the ones,
                can display less suppression under
                averaging.
                It is obvious
                that the rare events in the reaction
                represent the only reason
                for considering the integral quantities.

\section{
                Simulation of Data and Results
        }


                It is found necessary to study asymmetries
                of cross sections with respect to
                various planes in momentum space.
                The complicated
                form of
                the phenomenological amplitude
                makes it impossible to perform an
                analytic investigation
                of the polarization term
                of the matrix element
(\ref{SqAmpsRI}).
                In the absence
                of real experimental measurements,
                we perform theoretical simulations.
                Prior to discussion of their details
                given
                below,
                let us briefly consider the geometry
                of the existing
                CHAOS
                device
                which
                is capable to provide
                the necessary measurements
%
%
%
                (for more details see
\cite{SmithEtAl95,BonuttiEtAl93,BonuttiEtAl94}).

                The cylindrical dipole magnet,
                producing the vertical magnetic field,
                is the largest part of the CHAOS spectrometer.
                A polarized target is exposed
                to the horizontal pion beam.
                The target is inserted through the
                120~mm calibre hole
                along magnet's symmetry axis.
                So the target is placed at the center
                of the cylindrical space between
                the magnet poles,
                the pole diameter being
                950~mm.
                Four cylindrical chambers are surrounding
                the target:
                the most inner WC1 and WC2 are
                fast multi--wire proportional chambers,
                WC3 and WC4 are the drift chambers.

                The ring of gain stabilized
                counter telescopes
                constitutes the outside layer of detectors.
                These counters determine the vertical
                acceptance of
        $ \pm 7^{\circ} $.
                In the horizontal plane
                (CHAOS plane),
                there are
                deadened regions
                of WC3 and WC4
                in narrow angles
        ($ \approx 36^{\circ} $
                in total)
                where the beam enters and exits the device.
                This causes a difficulty for tracking
                and particle
                identification for some events.
                The angle and
                momentum value of
                an outgoing charged particle
                hitting only WC1 and WC2
                and
                missing WC3 and/or WC4
                are correlated
                because of the magnetic field present.
                We neglect this effect and simply set
                the horizontal acceptance to
        $ 360^{\circ} $.
                Apart from geometrical cuts,
                no extra factors
                such as efficiencies of registration,
                etc.
                are involved into our simulations
                for simplicity.

%


                Though some structures
                of polarized cross sections,
                like that of eqs.
(\ref{epspqskI}),
(\ref{epspqkRkI}),
                have no explicit dependence upon
                the relative orientation
                of the nucleon--spin vector
        $ s $
                and the beam
        $ k_{1} $,
                we consider two basic variants
                with respect to this orientation
                in the laboratory system.
                This is natural for the design
                of experimental devices
                and
                this is
                convenient for data simulations
                as well.
                So
        $ {\bf s} $
                is chosen to be orthogonal to the beam
        $ {\bf k}_{1} $
                and to the CHAOS plane in the first
                variant
        ($ {\bf s} {\bot} {\bf k}_{1} $)
                and
        $ {\bf s} $
                is chosen to be parallel to the beam
        $ {\bf k}_{1} $
                in the second variant
        ($ {\bf s} || {\bf k}_{1} $).

                The Monte--Carlo events for the reaction
                channel
        $ \pi^{-} p^{\uparrow} \to \pi^{-} \pi^{+} n $
                are generated at the beam momentum set to
        $ P_{\rm Lab} = 360 $~MeV/c.
                Few control runs are performed also for
                the rest channels
                at the same energy.
                We consider the
                azimuth angle in the
                orthogonal to
        ($ {\bf x} , {\bf y} $)
                plane through the vector
        $ {\bf y} $.
                The angle is counted out from the direction
        $ {\bf y} $.
                The bins for this angle
                are filled
                with
                the selection of events
                {\bf a})
                without geometrical restrictions
                and
                {\bf b})
                with restrictions of CHAOS geometry
                (CHAOS is hit by
                46 307 events
                from the requested amount
                of 2 000 000).

\begin{figure}[ht]
   \epsfxsize=16cm
   \epsfysize=16cm
   \centerline{\epsffile{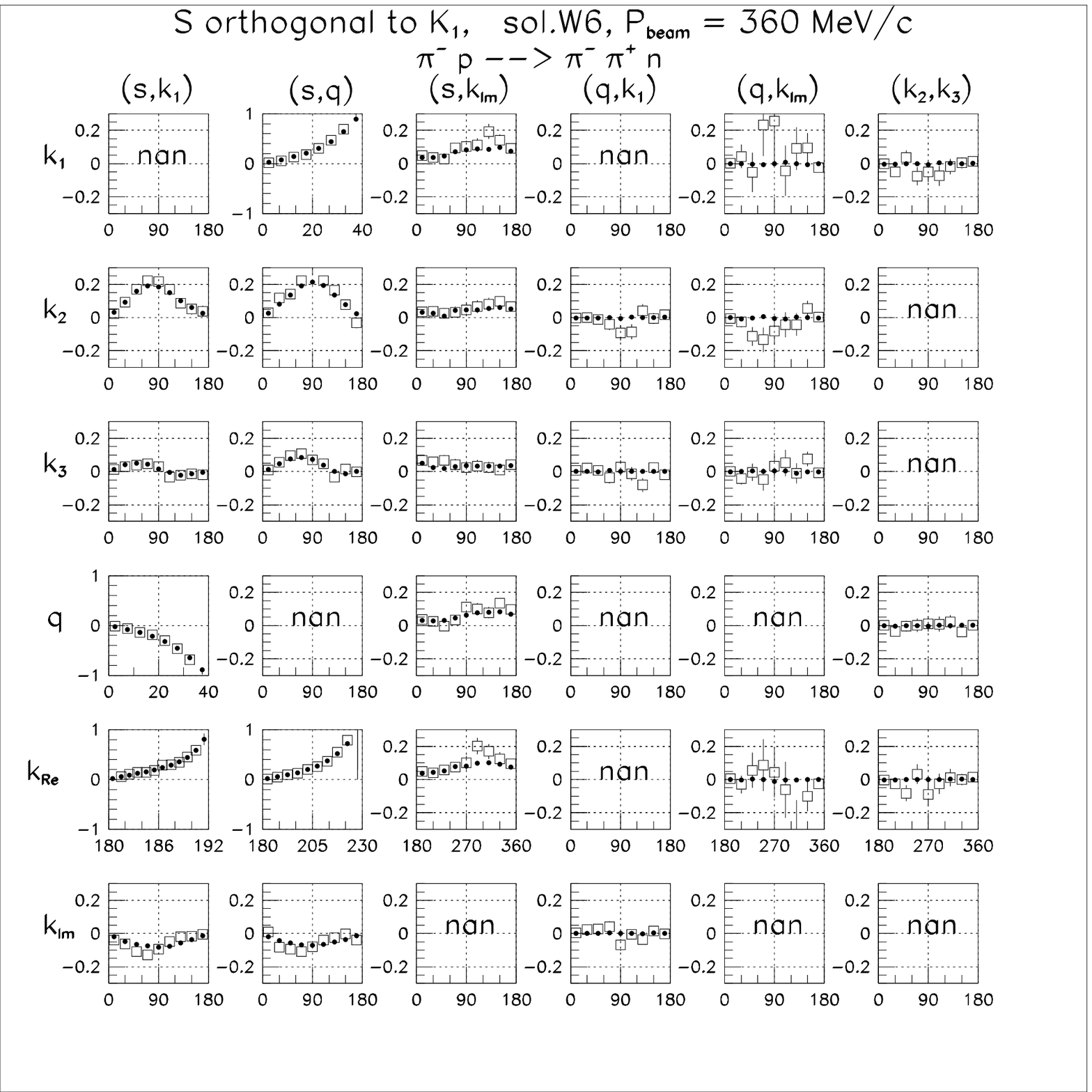}}
   \centerline{
               \parbox{15cm}{
                              \caption{
                                        \label{sTkw6}
                Asymmetries
                for the amplitude solw6
                at
        ($ {\bf s} {\bot} {\bf k}_{1} $)
                in the
        $ 4 \pi $--steradian geometry device
                (full dots)
                and in CHAOS
                (open squares).
                                      }
                             }
              }
\end{figure}

\begin{figure}[ht]
   \epsfxsize=16cm
   \epsfysize=16cm
   \centerline{\epsffile{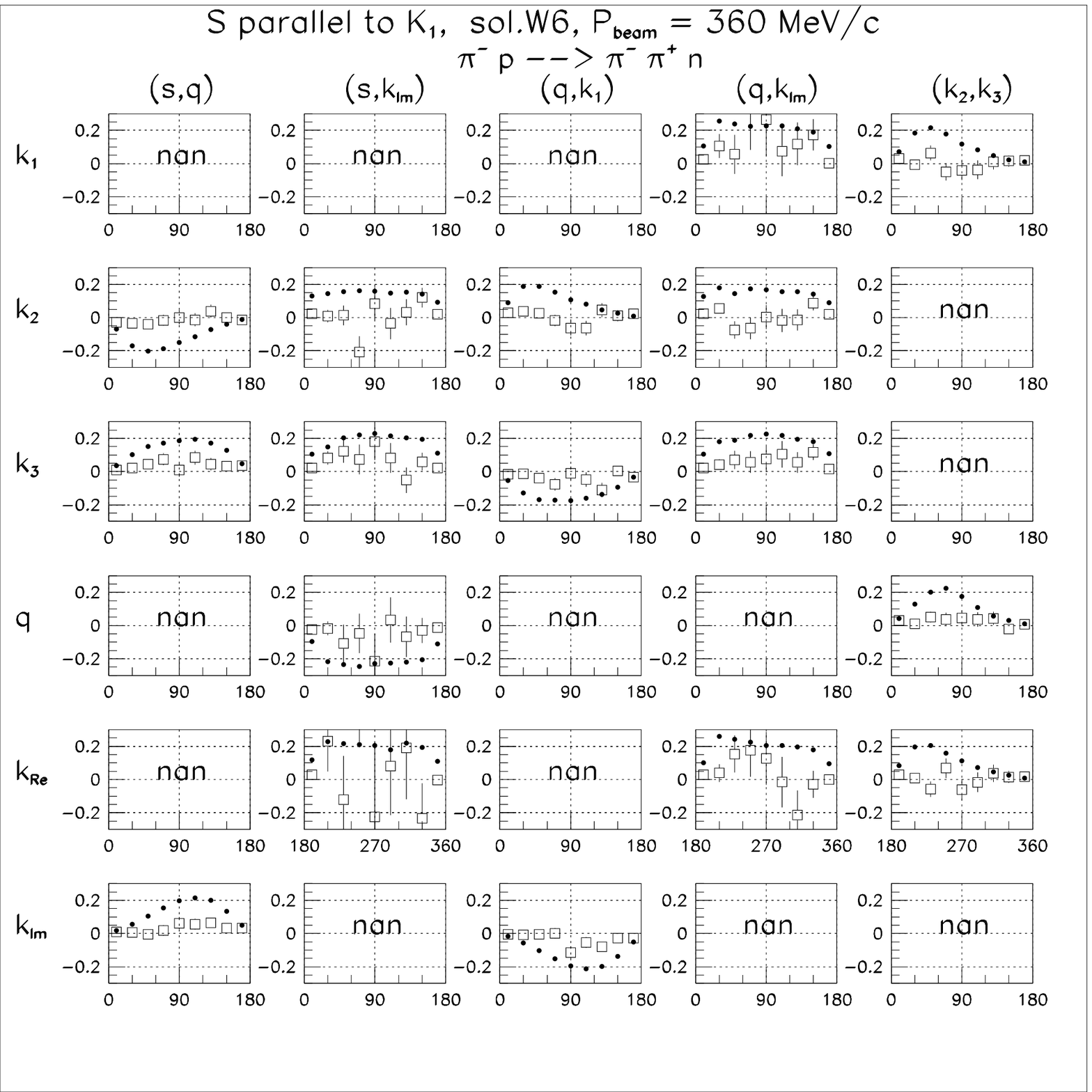}}
   \centerline{
               \parbox{15cm}{
                              \caption{
                                        \label{slkw6}
                Asymmetries
                for the amplitude solw6
                at
        ($ {\bf s} || {\bf k}_{1} $)
                in the
        $ 4 \pi $--steradian geometry device
                (full dots)
                and in CHAOS
                (open squares).
                                      }
                             }
              }
\end{figure}

\begin{figure}[ht]
   \epsfxsize=16cm
   \epsfysize=17cm
   \centerline{\epsffile{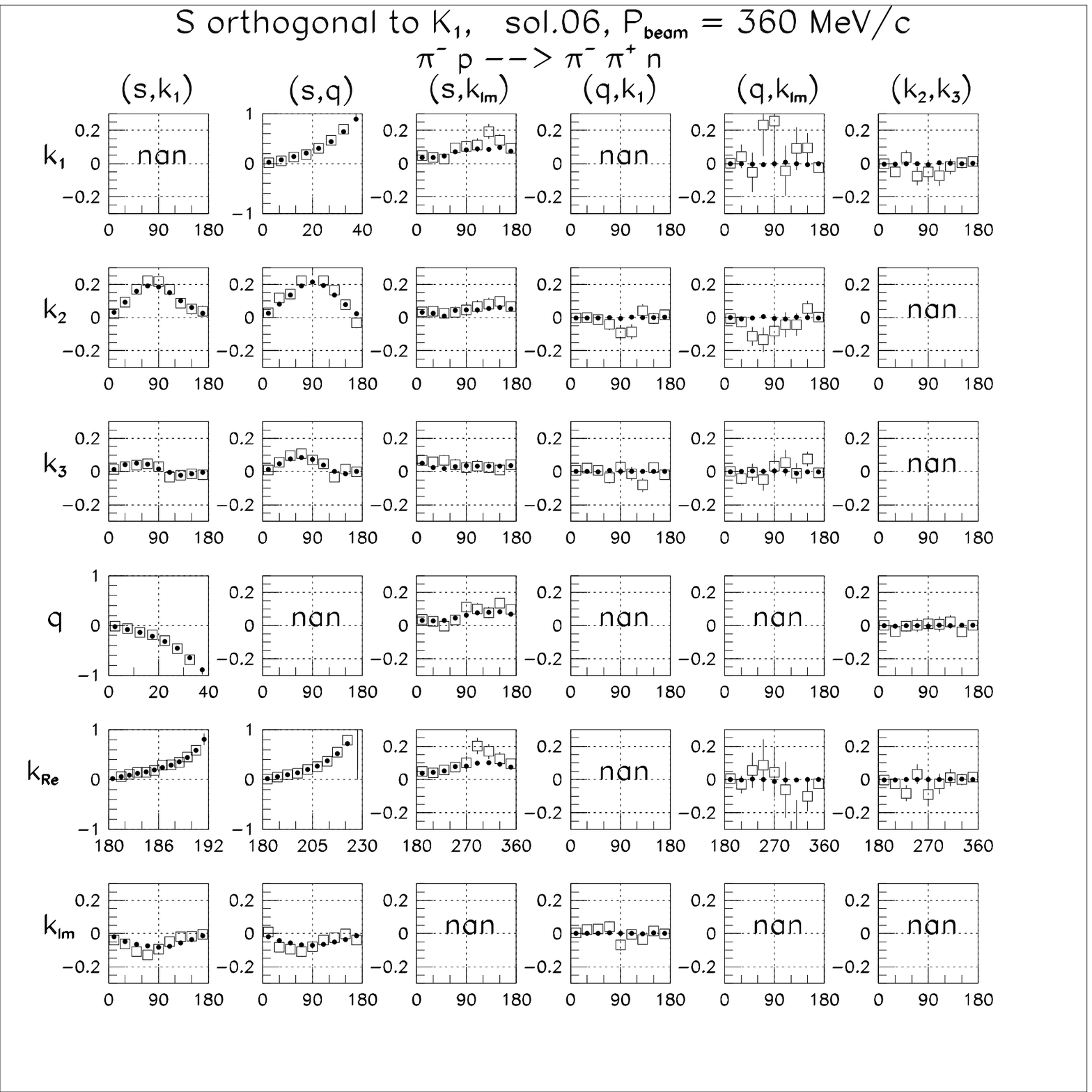}}
   \centerline{
               \parbox{15cm}{
                              \caption{
                                        \label{sTk06}
                Asymmetries
                for the amplitude sol06
                at
        ($ {\bf s} {\bot} {\bf k}_{1} $)
                in the
        $ 4 \pi $--steradian geometry device
                (full dots)
                and in CHAOS
                (open squares).
                                      }
                             }
              }
\end{figure}

\begin{figure}[ht]
   \epsfxsize=16cm
   \epsfysize=16cm
   \centerline{\epsffile{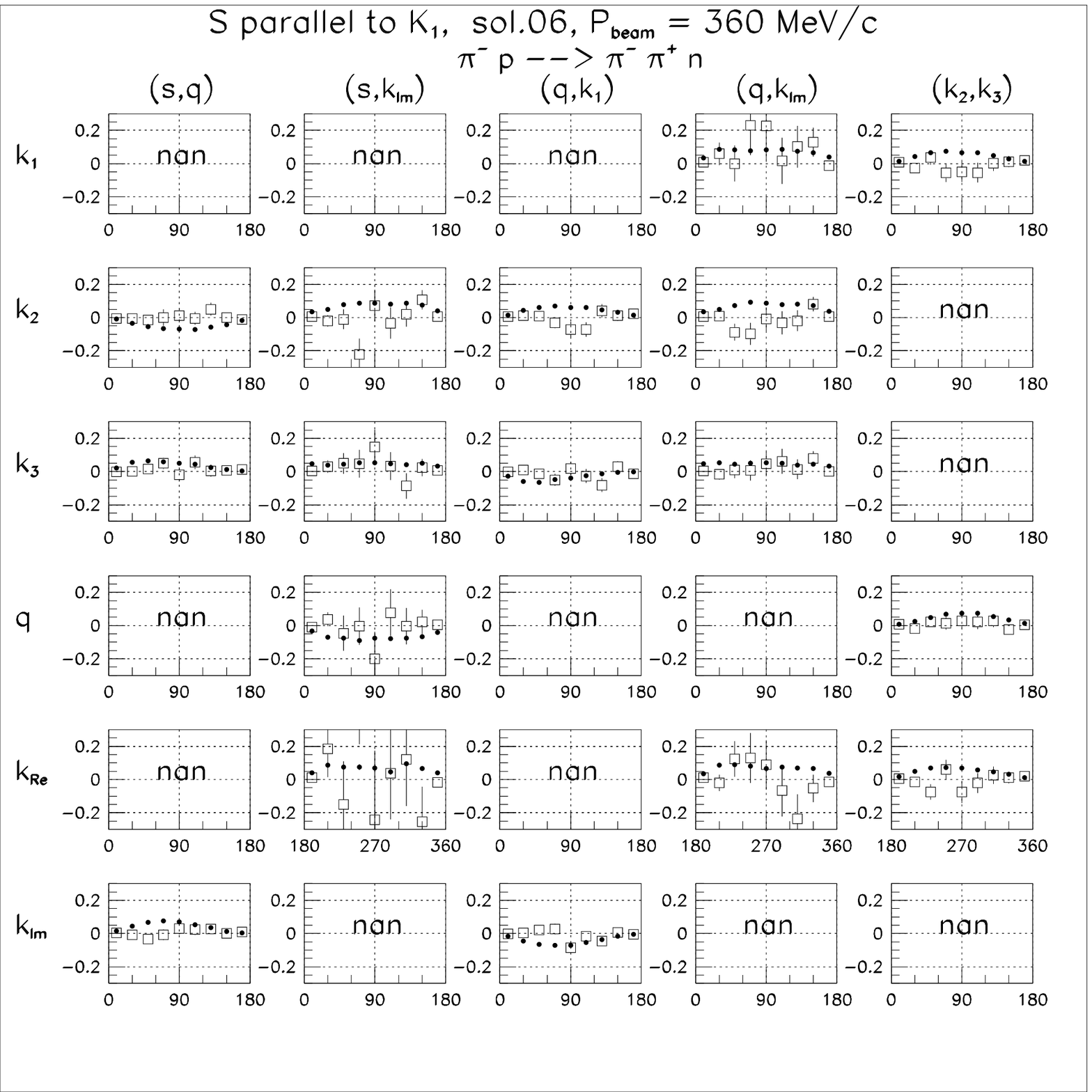}}
   \centerline{
               \parbox{15cm}{
                              \caption{
                                        \label{slk06}
                Asymmetries
                for the amplitude sol06
                at
        ($ {\bf s} || {\bf k}_{1} $)
                in the
        $ 4 \pi $--steradian geometry device
                (full dots)
                and in CHAOS
                (open squares).
                                      }
                             }
              }
\end{figure}

                The list of examined asymmetries
        $ A_{ ({\bf x, y}) } ( \varphi_{\bf z} ) $
                can be obtained from the headings
                of the figures
\ref{sTkw6}--\ref{slk06}.
                By obvious geometrical reasons
                the list for the variant
        ($ {\bf s} || {\bf k}_{1} $)
                is truncated.
                The list is far from being complete
                combinatorically.
                Nevertheless,
                it is sufficient to display the role
                of distinct structures and to demonstrate
                tight relations between
                quantities like
        $ A_{ ({\bf s, k}_{1}) } ( \varphi_{\bf q} ) $
                and
        $ A_{ ({\bf s, q}) } ( \varphi_{{\bf k}_{1}} ) $.

                Twelve
                amplitudes
                for
        $\pi N \to \pi \pi N$
                reactions,
                all of which but one
                being found
                as solutions
                of analysis
\cite{BolokhovPS98zs1}
                of unpolarized data,
                are used as the theoretical input.
                These solutions are practically equivalent
                by the
        $ \chi^{2} $
                criterion.
                Such properties of solutions like
        $ \chi^{2} $,
                errors of parameters, etc.
                are irrelevant for simulations.
                We show only specific values of
        $ \pi \pi $--scattering
                lengths
                in Table
\ref{solas}.
                The ordering of solutions in this Table
                is performed according to the value of
        $ a^{I=0}_{0} $.
                This reflects the role of OPE
                in the given amplitude:
                it is negligible for amplitudes from the
                bottom of Table
\ref{solas}.
                The discussed amplitudes can be split
                into two classes:
                the {\it physical} amplitudes,
                which support the sequence of signs
        $ \{ $ ``$ + $'', ``$ - $'', ``$ + $'' $ \} $
                for scattering lengths
        $ a^{I=0}_{0}, a^{I=2}_{0}, a^{I=1}_{1} $,
                and the rest amplitudes
                which we call
                {\it unphysical}.

                The amount of obtained data is too large
                to be displayed here.
                The following Table
\ref{solas}
                collects the largest values of
                integral asymmetries
                found for discussed amplitudes.
                Sometimes,
                a lower value is given
                if it is characterized by better accuracy.
                The given errors are only
                the statistical ones.
                This can help to estimate
                what number of experimental events
                is sufficient to detect
                the asymmetry in question.

\begin{table}
\begin{center}
\begin{tabular}{||c|c|c|c||c|c||c||}
\hline
  Ampl.
&
         $ a^{I=0}_{0} $
&
                       $ a^{I=2}_{0} $
&
                                     $ a^{I=1}_{1} $
&
        {($ {\bf s} {\bot} {\bf k}_{1} $)}
&
        {($ {\bf s} {\bot} {\bf k}_{1} $)}
&
        {($ {\bf s}  {||}  {\bf k}_{1} $)}
\\
&
&
&
&
        $ 4 \pi $
&
        CHAOS
&
        $ 4 \pi $
\\
\hline
  solw6 &   0.264 & -0.008 &  0.032 &
        $ 0.89(4)_{({\bf s},{\bf q})} ({\bf k}_{1})  $
&
             $ 0.91(7)_{({\bf s},{\bf q})} ({\bf k}_{1})  $
&
                $ 0.259(4)_{({\bf q},{\bf k}_{I})} ({\bf k}_{R})  $
\\
  sol05 &   0.193 &  0.074 & -0.014 &
        $ 0.251(9)_{({\bf s},{\bf q})} ({\bf k}_{R})  $
&
             $ 0.5(5)_{({\bf s},{\bf q})} ({\bf k}_{R})  $
&
                $ 0.203(4)_{({\bf q},{\bf k}_{I})} ({\bf k}_{R})  $
\\
  sol06 &   0.189 & -0.059 &  0.054 &
        $ -0.38(4)_{({\bf s},{\bf k}_{1})} ({\bf k}_{R})  $
&
             $ -0.40(6)_{({\bf s},{\bf k}_{1})} ({\bf k}_{R})  $
&
                $ 0.089(6)_{({\bf q},{\bf k}_{I})} ({\bf k}_{R})  $
\\
  sol04 &   0.175 &  0.048 & -0.084 &
        $ [-0.23(3)_{({\bf s},{\bf k}_{1})} ({\bf q})]  $
&
             $ [0.22(2)_{({\bf s},{\bf k}_{I})} ({\bf k}_{R})]  $
&
                $ 0.093(3)_{({\bf s},{\bf q})} ({\bf k}_{I})  $
\\
  sol10 &   0.172 & -0.043 &  0.050 &
        $ -0.41(3)_{({\bf s},{\bf k}_{1})} ({\bf q})  $
&
             $ 0.39(4)_{({\bf s},{\bf q})} ({\bf k}_{1})  $
&
                $ 0.133(3)_{({\bf q},{\bf k}_{I})} ({\bf k}_{R})  $
\\
  sol07 &   0.156 &  0.002 &  0.043 &
        $ 0.070(3)_{({\bf s},{\bf q})} ({\bf k}_{2})  $
&
             $ 0.12(4)_{({\bf s},{\bf k}_{I})} ({\bf k}_{R})  $
&
                $ 0.155(3)_{({\bf q},{\bf k}_{I})} ({\bf k}_{R})  $
\\
  sol02 &   0.105 &  0.036 & -0.027 &
        $ [-0.29(4)_{({\bf s},{\bf k}_{1})} ({\bf q})]  $
&
             $ [0.25(4)_{({\bf s},{\bf q})} ({\bf k}_{1})]  $
&
                $ 0.273(4)_{({\bf q},{\bf k}_{I})} ({\bf k}_{R})  $
\\
  sol09 &   0.077 & -0.006 &  0.036 &
        $ -0.52(5)_{({\bf s},{\bf q})} ({\bf k}_{1})  $
&
             $ 0.55(6)_{({\bf s},{\bf k}_{1})} ({\bf q})  $
&
                $ 0.284(4)_{({\bf q},{\bf k}_{I})} ({\bf k}_{R})  $
\\
  sol01 &   0.069 &  0.030 &  0.023 &
        $ -0.58(4)_{({\bf s},{\bf q})} ({\bf k}_{1})  $
&
             $ 0.61(6)_{({\bf s},{\bf k}_{1})} ({\bf q})  $
&
                $ 0.290(4)_{({\bf q},{\bf k}_{I})} ({\bf k}_{R})  $
\\
  sol03 &   0.069 & -0.057 &  0.045 &
        $ -0.29(3)_{({\bf s},{\bf k}_{1})} ({\bf q})  $
&
             $ 0.25(4)_{({\bf s},{\bf q})} ({\bf k}_{1})  $
&
                $ 0.265(8)_{({\bf s},{\bf k}_{I})} ({\bf k}_{2})  $
\\
  sol11 &   0.067 & -0.077 &  0.047 &
        $ 0.48(3)_{({\bf s},{\bf k}_{1})} ({\bf q})  $
&
             $ 0.52(5)_{({\bf s},{\bf k}_{1})} ({\bf q})  $
&
                $ 0.224(7)_{({\bf q},{\bf k}_{I})} ({\bf k}_{R})  $
\\
  sol12 &   0.011 & -0.008 & -0.030 &
        $ -0.281(3)_{({\bf s},{\bf q})} ({\bf k}_{3})  $
&
             $ 0.42(3)_{({\bf s},{\bf k}_{I})} ({\bf k}_{R})  $
&
                $ 0.381(4)_{({\bf q},{\bf k}_{I})} ({\bf k}_{R})  $
\\
\hline
\end{tabular}
\end{center}
\caption[orient.]{
                  The largest asymmetries
                for
        $ \pi^{-} p \to \pi^{-} \pi^{+} n $
                channel at
        $ P_{\rm Lab} = 360 $~MeV/c.
                The last--digit errors are given in brackets.
                The square brackets indicate the results
                for asymmetries developed under vanishing
                cross sections.
                 }
\label{solas}
\end{table}

\begin{figure}[ht]
   \epsfxsize=16cm
   \epsfysize=17cm
   \centerline{\epsffile{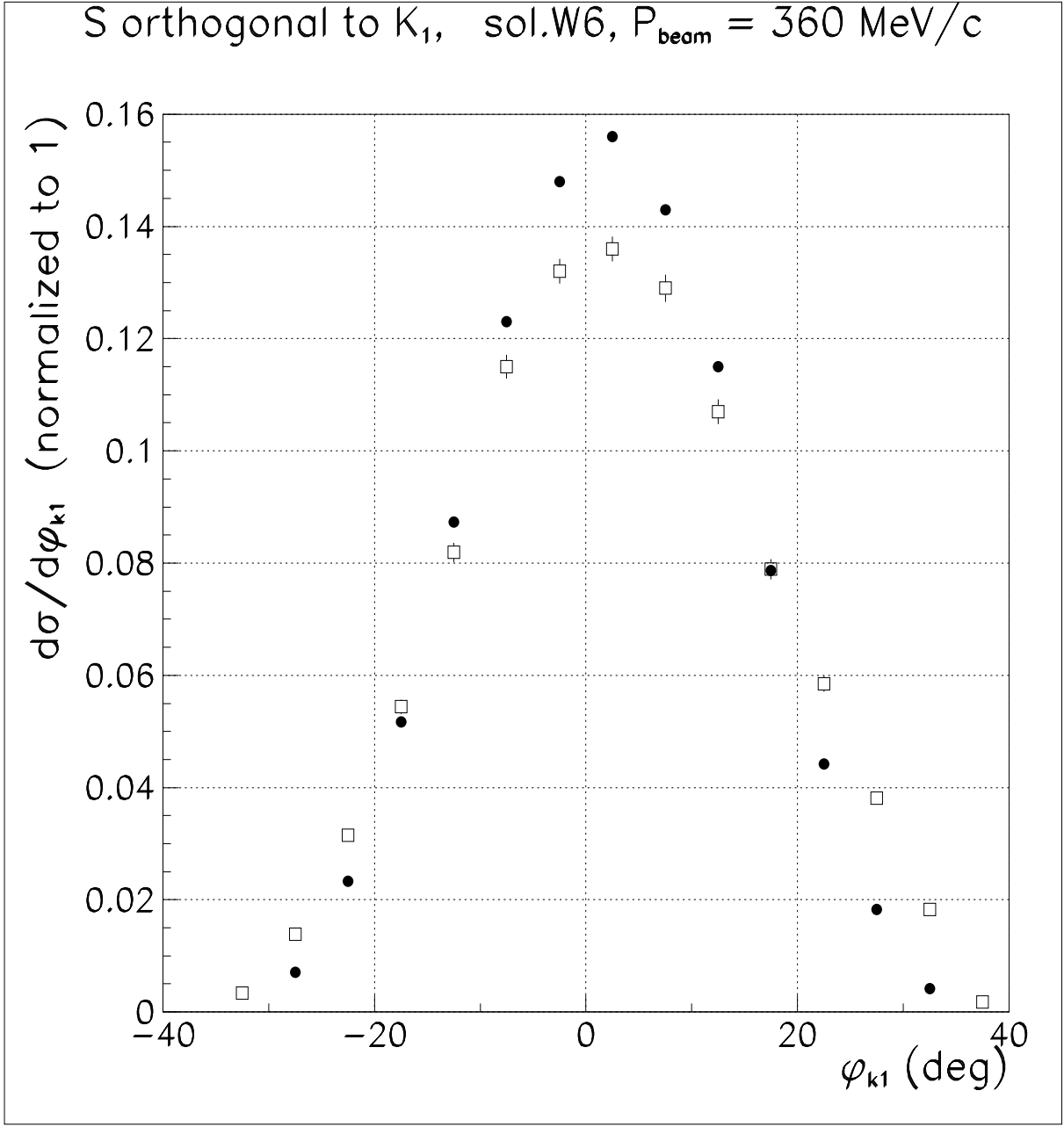}}
   \centerline{
               \parbox{15cm}{
                              \caption{
                                        \label{crsas}
                Distributions in the azymuth angle of
        $ {\bf k}_{1} $
                projected to the
                the plane through
        $ {\bf q} $
                and orthogonal to the plane
        ($ {\bf s}, {\bf q} $)
                for the amplitude solw6
                at
        ($ {\bf s} {\bot} {\bf k}_{1} $)
                in the
        $ 4 \pi $--steradian geometry device
                (full dots)
                and in CHAOS
                (open squares).
                                      }
                             }
              }
\end{figure}

\begin{figure}[ht]
   \epsfxsize=16cm
   \epsfysize=16cm
   \centerline{\epsffile{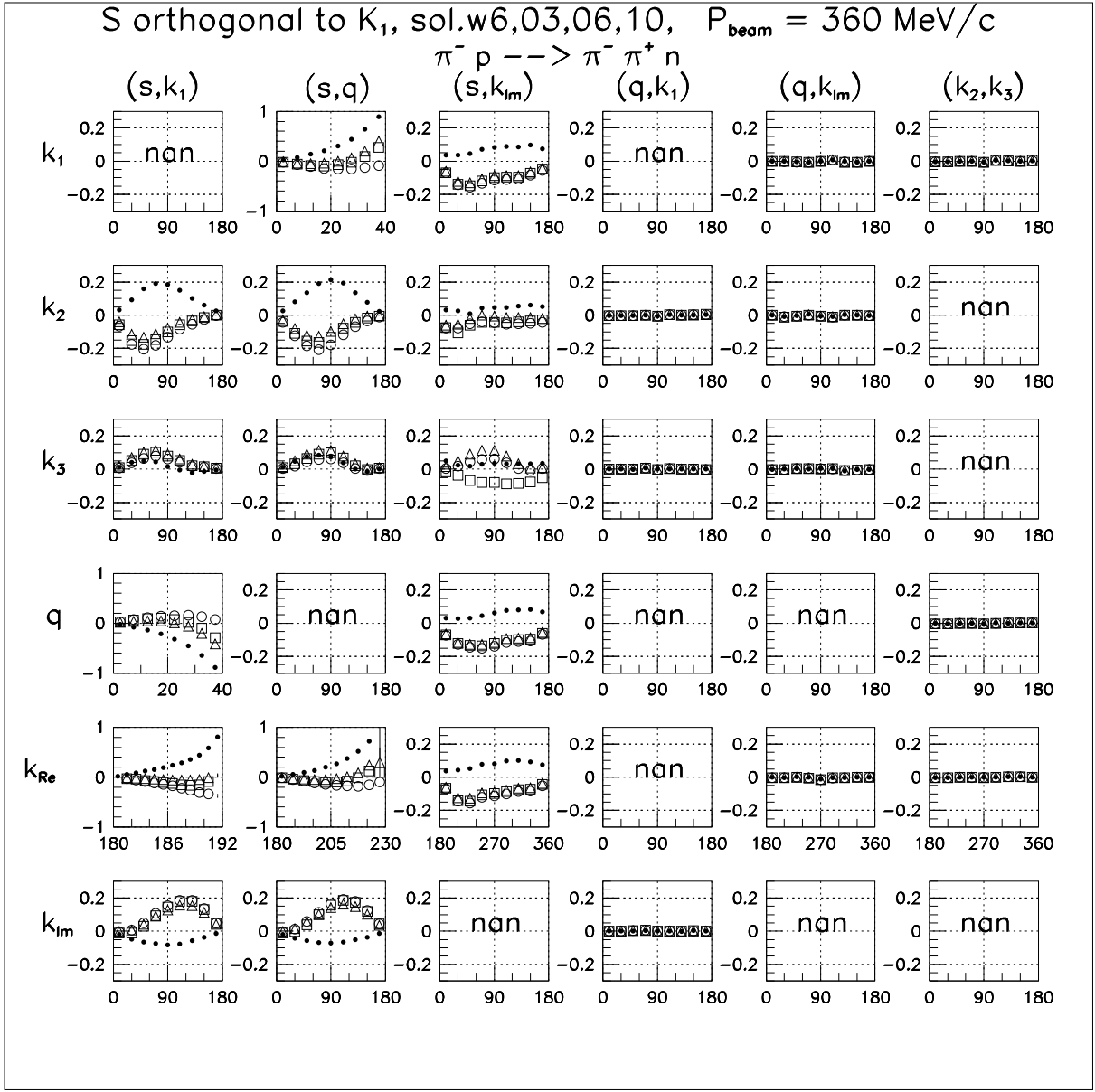}}
   \centerline{
               \parbox{15cm}{
                              \caption{
                                        \label{sTk060310w6}
                Asymmetries
                for the amplitudes
                sol06
                (open circles),
                sol03
                (open squares),
                sol10
                (open triangles),
                solw6
                (full dots)
                at
        ($ {\bf s} {\bot} {\bf k}_{1} $)
                in the
        $ 4 \pi $--steradian geometry device.
                                      }
                             }
              }
\end{figure}

                All data for asymmetries were also
                represented in the graphical form.
                Only few of them are shown in figs.
\ref{sTkw6}--\ref{sTk060310w6}
                for illustrative purpose.
                For example,
                a comparison of fig.
\ref{sTkw6}
                with fig.
\ref{slkw6}
                and fig.
\ref{sTk06}
                with fig.
\ref{slk06}
                helps to make conclusions on the role
                of initial--spin orientation
                and on different properties
                of CHAOS selections.
                Comparing figs.
\ref{sTkw6},
\ref{slkw6}
                with figs.
\ref{sTk06},
\ref{slk06},
                one finds
                quite individual polarization phenomena
                for different solutions.


                The collection of pictures is found to have
                a striking property:
                all solutions,
                being indistinguishable by
        $ \chi^{2} $
                in the course of analysis
\cite{BolokhovPS98zs1},
                appear to be different.

                The examination of pictures shows also
                that the asymmetries
             $ A_{({\bf s},{\bf q})} (\varphi_{{\bf k}_{1}})  $,
             $ A_{({\bf s},{\bf q})} (\varphi_{{\bf k}_{R}})  $,
             $ A_{({\bf s},{\bf k}_{1})} (\varphi_{\bf q})  $
                are characteristic of the OPE mechanism.
                These asymmetries become smaller
                and other asymmetries start to appear
                when one is going from solutions from the top
                of Table
\ref{solas}
                to solutions from the bottom,
                i.e. ``switching off'' the OPE contribution.

                Another important feature of the obtained
                data
                is related to unresolved ambiguity between
                the physical and unphysical solutions.
                It is found that the latter develop
                smaller asymmetries.
                Generally,
                the asymmetries for unphysical solutions
                are more difficult to detect,
                since they are peaking in narrow angles
                characterized by low cross sections.
                In contrast,
                when asymmetries for physical amplitudes
                reach the maximal values,
                the curves are gently sloping.
                The found absolute maximum
        $ A \sim 1 $
                corresponds usually to directions with
                small cross sections
                (see distributions at
        $ \varphi = 40^{\circ} $
                given in
                fig.
\ref{crsas};
                the corresponding asymmetry
        $ A_{({\bf s},{\bf q})} (\varphi_{{\bf k}_{1}})  $
                can be found in fig.
\ref{sTkw6}).
                There are enough statistics for the
                nearby angles to detect the relatively high
                value of such asymmetry,
                for example, at
        $ \varphi \sim 20^{\circ} $,
                see fig.
\ref{crsas}.

                When putting aside the last amplitudes from
                Table
\ref{solas}
                and splitting the rest into physical
                and unphysical groups,
                a regular behaviour of pictures with the
                parameter
        $ a^{I=2}_{0} $
                can be found in both groups.
                This is demonstrated by fig.
\ref{sTk060310w6},
                where the asymmetries for physical
                amplitudes
                sol06, sol03, sol10 and solw6
                are shown at
        ($ {\bf s} \bot {\bf k}_{1} $).
                The almost smooth transformation of
                one picture into another is clearly seen
                for asymmetries,
                which are relevant to OPE.
                This regularity and the absence of the
                same regularity for
        $ a^{I=0}_{0} $
                variation
                can be interpreted
                as an indirect evidence in favour of
                smaller perturbations
                by isobar contributions
                to isospin
        $ {I=2} $
                amplitude.
                However,
                poor asymmetries from OPE and rich the rest ones
                obtained for
        $ \pi^{+} p^{\uparrow} \to \pi^{+} \pi^{+} n $
                channel
                do not support this.

                These conclusions are valid for measurements
                with
        ($ {\bf s} \bot {\bf k}_{1} $)
                in devices with
        $ 4\pi $--steradian
                geometry and in the CHAOS device as well.
                Moreover,
                the CHAOS geometry selects events displaying
                larger asymmetries,
                though at the price of lower statistics.

                The asymmetries for the setup
        ($ {\bf s} || {\bf k}_{1} $)
                are rich and informative for the
        $ 4\pi $--steradian
                geometry of a hypothetical device.
                Here, the asymmetries
        $ A_{({\bf q},{\bf k}_{1})} (\varphi_{\bf z})  $,
        $ A_{({\bf q},{\bf k}_{I})} (\varphi_{\bf z})  $,
        $ A_{({\bf k}_{2},{\bf k}_{3})} (\varphi_{\bf z}) $
        ($ {\bf z} = {\bf k}_{1}, {\bf k}_{2}, {\bf k}_{3},
                     {\bf k}_{R},  {\bf k}_{I}, {\bf q}  $),
                all of which are almost flat for
        ($ {\bf s} \bot {\bf k}_{1} $)
                setup,
                are looking much more vivid
                (cf. figs.
\ref{sTkw6}
                and
\ref{slkw6}).
                According to criteria of sect.
\ref{xsectPT},
                the ``switching off'' effect of
        $ ({\bf s} \cdot {\bf k}_{R}) \sim $
        $ - ({\bf s} \cdot  {\bf k}_{1}) = 0 $
                must be solely due to
                non--OPE mechanisms in the test amplitudes.
                These are indeed present
                in all discussed solutions.

                When projected to CHAOS,
                all examined asymmetries for
        ($ {\bf s} || {\bf k}_{1} $)
                appear
                to be consistent with zero.
                This is not so surprising,
                since,
                for the
        ($ {\bf s} || {\bf k}_{1} $)
                setup,
                practically
                all interesting events happen
                in the plane
                which is orthogonal to the beam
        $ {\bf k}_{1} $.
                The most part of such events avoids
                CHAOS chambers.
                Though the cross sections themselves
                are found to be sensitive
                to the tested
                amplitudes
                in the forward and backward cones,
                it is difficult to evaluate the importance
                of such data.
                One can recall
                that the above critical plane is entirely
                accepted by the wire--chamber space of the
                design for the AMPIR spectrometer
\cite{KurepinEtAl92}.
                This remarkable complementarity
                of CHAOS and AMPIR devices
                makes a promise
                for
                exhaustive investigations of
                polarization effects in
        $ \pi N \to \pi \pi N $
                reactions at low energies.

                We finish the discussion of results
                by reminding
                that there are simplifications
                in the procedure.
                The incomplete polarization in the real
                experiment can decrease the absolute values
                of the shown asymmetries by few percent.
                The real--device efficiency
                and the reduced experimental statistics
                enlarge errors.
                The number of generated events
                in simulations
        ($ \sim 4 \times 10^{4} $
                hitting CHAOS)
                represents the limit
                ever attainable experimentally.
                Provided no confident result is obtained,
                the problem of polarization phenomena
                below isobar threshold
        $ P_{\rm Lab} < 500 $~MeV/c
                would be closed.
                The rich picture of effects
                displayed by almost every amplitude at
        $ P_{\rm Lab} = 360 $~MeV/c
                is far above cautious expectations.
                The magnitude of statistical errors
                given in Table
\ref{solas}
                demonstrates that the number of experimental
                events
                which is necessary to detect the discussed
                phenomena
                and discriminate
                between competing contributions,
                can be decreased by an order of magnitude.
                It must be noted that reaction channel
        $ \pi^{-} p^{\uparrow} \to \pi^{-} \pi^{+} n $
                is not the best one in respect to integral
                asymmetries examined in this section.
                Simulations with some selected amplitudes
                show the following order of preference:
        $ \pi^{+} p^{\uparrow} \to \pi^{+} \pi^{0} p $,
        $ \pi^{-} p^{\uparrow} \to \pi^{-} \pi^{0} p $,
        $ \pi^{-} p^{\uparrow} \to \pi^{-} \pi^{+} n $,
        $ \pi^{+} p^{\uparrow} \to \pi^{+} \pi^{+} n $,
        $ \pi^{-} p^{\uparrow} \to \pi^{0} \pi^{0} n $.
                This means that the neutral channel
                requires full--kinematics measurements
                for detecting asymmetries in distinct
                regions of the entire phase space.


\section{ Conclusions }


                The main achievement of the present paper
                is the demonstration of striking efficiency
                of the polarization measurements for
        $ \pi N \to \pi \pi N $
                reactions
                in the very close to threshold energy region.
                Such measurements are feasible
                with the use of CHAOS spectrometer
                right now.
                This is shown
                in the framework of the standard formalism
                adjusted to the canonical form of the
        $ \pi N \to \pi \pi N $
                amplitude
\cite{BolokhovVS88,BolokhovVS91}.

                The motivation for such experiments follows
                from disappointing difficulties
                encountered within frameworks of all
                known methods for the analysis of
                low--energy
        $ \pi N \to \pi \pi N $
                data.
                The deep reason of difficulties in
                the interpretation of the
        $ \pi N \to \pi \pi N $
                results
                is related to the very nature of the
                unpolarized data
                which cannot help to discriminate
                between the
        $ t $--channel
                mechanism of OPE and isobar exchanges.

                The competition between OPE and the
                rest mechanisms of
        $ \pi N \to \pi \pi N $
                reaction,
                preventing from accurate determination of
        $ \pi \pi $
                interaction
                with the help of unpolarized data,
                at the same time gives rise to extremely
                rich polarization effects
                within half--width isobar region.

                The effects are found to be sensitive
                as to the OPE parameters in question
                as well as
                to details of isobar interactions.
                All equivalent solutions of the paper
\cite{BolokhovPS98zs1}
                appear to be different
                in the asymmetry picture.
                Therefore,
                any project of determination of parameters
                of
        $ \pi \pi $
                interaction
                with the help of
        $ \pi N \to \pi \pi N $
                data
                must assume the polarization measurements.
                The yield for the decisive
        $\pi N \to \pi \pi N$
                analysis
		must combine both unpolarized data
		and polarization information.

                The application of
                results of the present paper
                is straightforward
                within the method
        {\bf 3}
                mentioned in Introduction.
                It is simple to find
                that all asymmetries vanish in the
                extrapolation points specific to methods
        {\bf 1}
                and
        {\bf 2}.
                This can help to estimate a part
                of the theoretical error characteristic
                of a method.
                Indeed,
                provided
                the data
                of polarized--target experiments
                are collected separately
                from right
                and left semi--spheres with respect to
        ($ {\bf s}, {\bf k}_{1} $),
        ($ {\bf s}, {\bf q} $)
                or
        ($ {\bf q}, {\bf k}_{1} $)
                planes,
                the estimate of error is obtained by
                the independent extrapolations.

\section{ Acknowledgments }

                AAB and SGS thank
                Russian Federation for Basic Research
                for support in terms
                of grant N 95-02-05574a.
                We are greatful to
                G.A.~Feofilov and
                D.~Po\v{c}ani\'{c}
                for remarks and to members of
                CHAOS collaboration
                        P.~Amaudruz,
                        F.~Bonutti,
                        J.~Brack,
                        P.~Camerini,
                        E.~Fragiacomo,
                        N.~Grion,
                        G.~Hofman,
                        R.R.~Johnson,
                        S.~McFarland,
                        M.~Kermani,
                        R.~Rui,
                        M.~Sevior,
                        G.R.~Smith,
                        R.~Tacik
                for various help and discussions.


%

\newpage

\vspace{1.5cm}

\newpage

\vspace{1.5cm}

\newpage

\vspace{1.5cm}

\newpage

\vspace{1.5cm}

\newpage

\vspace{1.5cm}

\newpage

\vspace{1.5cm}

\end{document}